\documentclass[prc,preprint,superscriptaddress,showpacs,nofootinbib,tightenlines,amsmath]{revtex4}
\usepackage{graphics}
\usepackage{epsfig}
\usepackage{slashed}
\usepackage{pbox, calc}
\usepackage{nicefrac}

\usepackage{xparse}
\usepackage{tikz}
\usepackage{pgfpages}
\usetikzlibrary{calc,backgrounds,trees,positioning,arrows,chains,shapes.geometric,%
    decorations.pathreplacing,decorations.pathmorphing,decorations.markings,shapes,%
    matrix,shapes.symbols}
\tikzstyle{every picture}+=[remember picture] \tikzstyle{na} =
[baseline=-.5ex] \tikzstyle{background grid}=[draw,
black!50,step=.5cm]
\usetikzlibrary{calc,shapes,arrows,automata,backgrounds,calendar}
\usetikzlibrary{intersections}
\usetikzlibrary{matrix}
\usetikzlibrary{trees}
\usetikzlibrary{decorations.pathmorphing}
\usetikzlibrary{decorations.pathreplacing}
\usetikzlibrary{decorations.markings}
\usetikzlibrary{3d}
\usetikzlibrary{external}
\pgfdeclaredecoration{complete sines}{initial} {
  \state{initial}[
    width=+0pt,
    next state=sine,
    persistent precomputation={\pgfmathsetmacro\matchinglength{
        \pgfdecoratedinputsegmentlength / int(\pgfdecoratedinputsegmentlength/\pgfdecorationsegmentlength)}
        \setlength{\pgfdecorationsegmentlength}{\matchinglength pt}
    }] {}
  \state{sine}[width=\pgfdecorationsegmentlength]{
    \pgfpathsine{\pgfpoint{0.25\pgfdecorationsegmentlength}{0.5\pgfdecorationsegmentamplitude}}
    \pgfpathcosine{\pgfpoint{0.25\pgfdecorationsegmentlength}{-0.5\pgfdecorationsegmentamplitude}}
    \pgfpathsine{\pgfpoint{0.25\pgfdecorationsegmentlength}{-0.5\pgfdecorationsegmentamplitude}}
    \pgfpathcosine{\pgfpoint{0.25\pgfdecorationsegmentlength}{0.5\pgfdecorationsegmentamplitude}}
  }
  \state{final}{}
}

\tikzset{
  photon/.style=  {
                    decorate,
                    solid,
                    decoration={complete sines, amplitude=1mm, segment length=1.75mm, post length=0}
                  },
  rho/.style= {
                solid
              },
  rhoarrow/.style=  {
                  solid,
                  postaction={decorate},
                  decoration={markings, mark=at position .6 with {\arrow{>}}}
                }
}
 \tikzset{->-/.style={decoration={
  markings,
  mark=at position #1 with {\arrow{>}}},postaction={decorate}}}
\tikzset{-<-/.style={decoration={
  markings,
  mark=at position #1 with {\arrow{<}}},postaction={decorate}}}

\begin{document}
\title{Interaction of the vector-meson octet with the baryon octet
in effective field theory}
\author{Y. \"Unal}
\affiliation{PRISMA Cluster of Excellence, Institut f\"ur
Kernphysik, Johannes Gutenberg-Universit\"at Mainz, D-55099 Mainz,
Germany}
\affiliation{Physics Department, \c{C}anakkale  Onsekiz Mart University, 17100 \c{C}anakkale, Turkey}
\author{A. K\"u\c{c}\"ukarslan}
\affiliation{Physics Department, \c{C}anakkale  Onsekiz Mart University, 17100 \c{C}anakkale, Turkey}
\author{S.~Scherer}
\affiliation{PRISMA Cluster of Excellence, Institut f\"ur
Kernphysik, Johannes Gutenberg-Universit\"at Mainz, D-55099 Mainz,
Germany}
\date{October 8, 2015}
\preprint{MITP/15-040}
\begin{abstract}
   We analyze the constraint structure of the interaction of vector mesons with baryons using the classical Dirac constraint analysis.
   We show that the standard interaction in terms of two independent SU(3) structures is consistent at the classical level.
   We then require the self-consistency condition of the interacting system in terms of perturbative renormalizability
to obtain relations for the renormalized coupling constants at the one-loop level.
   As a result we find a universal interaction with one coupling constant which is the same as in the massive
Yang-Mills Lagrangian of the vector-meson sector.
\end{abstract}
\maketitle

\section{Introduction}
   The ground-state baryon octet as well as the vector-meson octet played a vital role in shaping
our understanding of the symmetries of the strong interactions
(for an overview, see, e.g., Ref.\ \cite{GellMann:1964xy}).
   According to Coleman's theorem \cite{Coleman:1966}, the multiplet structure of the light
hadrons is related to an approximate SU(3) symmetry of the ground state of QCD.
   In fact, in the limit of massless up, down, and strange quarks, the
QCD Lagrangian exhibits a chiral $\text{SU(3)}_L\times\text{SU(3)}_R$ symmetry,
which is assumed to be dynamically broken down to $\text{SU(3)}_V$ in the ground state.
   As a result of this mechanism one expects the appearance of eight massless Goldstone bosons
\cite{Nambu:1961tp,Goldstone:1961eq,Goldstone:1962es}, which are identified with
the members of the pseudoscalar meson octet.
   The masses of the pseudoscalars in the real world are attributed to an explicit
chiral symmetry breaking due to the finite quark masses.
   The masses of hadrons other than the Goldstone bosons stay finite in
the chiral limit.

   Symmetry considerations not only affect the spectrum of QCD
but also put constraints on the interaction among hadrons.
   The dynamics of hadrons may be described in terms of an effective field theory \cite{Weinberg:1978kz}.
   To that end, one considers the most general Lagrangian compatible with the symmetries
of the underlying theory.
   Given a power-counting scheme, one may then calculate observables in terms
of perturbation theory or, alternatively, by applying non-perturbative methods such as
solving integral equations.
   While the interaction of the pseudoscalar octet $(\pi,K,\eta)$ with the baryon
octet is largely constrained by spontaneous symmetry breaking (see, e.g., Ref.\ \cite{Scherer:2012zzd}),
this is not the case for the coupling of the vector-meson octet to the baryon octet.
   Moreover, when describing the dynamics of vector mesons in a Lagrangian framework, one
inevitably faces the following challenge.
   Effective Lagrangians for vector particles (spin $S=1$, parity
$P=-1$) are constructed with Lorentz four-vector fields $V^\mu$ or
anti-symmetric tensor fields $W^{\mu\nu}=-W^{\nu\mu}$ with four and six
independent fields, respectively (see, e.g., Refs.~\cite{Ecker:1989yg,Birse:1996hd}).
   Therefore, one imposes constraints which, for an interacting theory,
may lead to relations among the coupling constants of the Lagrangian.
   For example, by applying a Dirac constraint analysis \cite{Dirac:2001}
to the interaction of the pion triplet with the Delta quadruplet, it was shown in Ref.~\cite{Wies:2006rv} that
the number of independent coupling constants reduces from three at the Lagrangian
level to a single coupling.
   Additional constraints beyond the consistency at the classical level may be obtained
if we require the theory to be perturbatively renormalizable
in the sense of effective field theory \cite{Weinberg:mt}.
   An investigation of this type for the pure vector-meson sector results in
a massive Yang-Mills theory \cite{Djukanovic:2010tb,Neiser:2011,Bijnens:2014fya}.
   {\it All} case studies found a reduction in the number of parameters
which seemed to be independent from the point of view of constructing
the most general Lagrangian.
   This is of particular importance when working with purely
phenomenological Lagrangians, because one is likely to introduce more
structures, and thus seemingly free parameters, than allowed by
a self-consistent treatment.

   In the present article, we want to focus on the lowest-order effective Lagrangian
for the interaction of the vector-meson octet with the ground-state baryon octet.
   For that purpose, in Sec.~\ref{section_review}, we will summarize the idea
of the Dirac constraint analysis.
   In Sec.~\ref{section_lagrangian}, we define the relevant Lagrangians and then
apply the Dirac constraint analysis in Sec.\ \ref{section_classical_constraint_analysis}.
   In Sec.~\ref{section_renormalizability}, we investigate the constraints
resulting from renormalizability in the sense of effective field theory
at the one-loop level.
   Our results are summarized in Sec.~\ref{section_summary}.
   Some technical details are relegated to the appendices.

\section{Review of the Dirac constraint analysis}
\label{section_review}

   A common procedure for the quantization of a classical system with given symmetries
is to first construct the Lagrangian of the system, which is assumed to be invariant
under the corresponding transformation of the dynamical variables,
and then to perform the transition to the Hamiltonian in terms of a Legendre transformation.
   On the one hand, the Lagrangian formalism is suitable for satisfying Lorentz invariance
and other symmetries, on the other hand, the Hamiltonian formalism is needed to
calculate the $S$ matrix \cite{Weinberg:mt}.
   For a system including constraints, we perform the transition from the Lagrangian to the Hamiltonian
by applying Dirac's constraint analysis to be discussed below \cite{Dirac:2001,Gitman:1990,Henneaux:1992}.
   The quantization of the constrained system is performed using path-integral methods
\cite{Gitman:1990,Henneaux:1992,Djukanovic:2010tb}.

   In the following, we will summarize Dirac's constraint analysis in terms of
a classical system with a finite number $N$ of degrees of freedom (DOF).
   To start with, we consider a Lagrange function $L(q,\dot{q})$ which depends on $N$ coordinates $q_i$ and
the corresponding velocities $\dot{q}_i=\frac{dq_i}{dt}$, collectively denoted by $q$ and $\dot{q}$, respectively.
   We assume that $L$ does not explicitly depend on time and that the $\dot{q}_i$ appear in monomials
of maximal degree 2 in $L$:
\begin{equation}
L(q,\dot{q})=\frac{1}{2}A_{ij}(q)\dot{q}_i\dot{q}_j+b_i(q)\dot{q}_i+c(q),
\end{equation}
where $A_{ij}=A_{ji}$, i.e., $A=(A_{ij})$ is a symmetric $N\times N$ matrix, $A=A^T$.
   To perform the transition to the Hamilton formalism,
one needs to introduce the momenta $p_i$ conjugate to the coordinates $q_i$,
\begin{equation}
p_i=\frac{\partial L(q,\dot{q})}{\partial \dot{q}_i}=A_{ij}(q)\dot{q}_j+b_i(q),
\label{pi}
\end{equation}
or
\begin{equation}
\label{pialternative}
p(q,\dot{q})=A(q)\dot{q}+b(q).
\end{equation}
   Because the Hamiltonian is a function of $(q,p)$, one needs to be able to invert Eq.~(\ref{pi}) to go over from the set of dynamical
variables $(q,\dot{q})$ to $(q,p)$.
   To uniquely solve Eq.~(\ref{pialternative}) for the velocities, the existence
of the inverse matrix $A^{-1}$ is required,
\begin{equation}
\dot{q}=A^{-1}(p-b),
\end{equation}
where
\begin{equation}
A_{ij}=\frac{\partial}{\partial\dot{q}_j}p_i=\frac{\partial}{\partial{\dot{q}_j}}\frac{\partial L}{\partial\dot{q}_i}
=\frac{\partial^2 L}{\partial\dot{q}_j\partial\dot{q_i}}=
\frac{\partial^2 L}{\partial\dot{q}_i\partial\dot{q_j}}=
A_{ji}.
\end{equation}
   In other words, for a unique description of the velocities in terms of the momenta,
the Jacobian matrix $\partial(q,p)/\partial(q,\dot{q})$ cannot be singular, i.e.,
\begin{eqnarray}
\text{det} \left(\frac{\partial^2 L}{\partial \dot{q}_i \partial \dot{q}_{j}}\right)
\label{d}
\end{eqnarray}
cannot vanish.
   However, in the case that the determinant vanishes, the theory is singular and one cannot pass
from the Lagrange function to the Hamiltonian formulation in the standard manner.
   In this case, we make use of a method originally proposed by Dirac \cite{Dirac:2001}.

   In a singular system, we are not able to determine all velocities as functions of the coordinates and the independent momenta.
   Let the unsolvable $\dot q_i$ be the first $M$ velocities
$\dot q_1,\ldots, \dot q_M$.
   The so-called {\it primary} constraints occur as follows.
   The Lagrange function $L$ can be written as
\begin{equation}
L(q,\dot q)=\sum_{i=1}^M F_i(q) \dot q_i+G(q,\dot
q_{M+1},\ldots,\dot q_N),
\end{equation}
from which we obtain as the canonical momenta
\begin{equation}
p_i=\left\{\begin{array}{ll}F_i(q)&\textnormal{for } i=1,\ldots,M,\\
\frac{\partial G(q,\dot q_{M+1},\ldots,\dot q_N)}{\partial\dot q_i}&\textnormal{for
}i=M+1,\ldots,N.
\end{array}\right.
\label{kanonimp'}
\end{equation}
   The first part of Eq.\ (\ref{kanonimp'}) can be reexpressed
in terms of the relations
\begin{equation}
\phi_i(q,p)=p_i-F_i(q)\approx 0,\quad i=1,\ldots,M,
\end{equation}
which are referred to as the primary constraints.
   Here, $\phi_i\approx 0$ denotes a weak equation in Dirac's
sense, namely that one must not use one of these constraints before
working out a Poisson bracket \cite{Dirac:2001}.
   Using
\begin{equation}
H(q,p)=\sum_{i=1}^N p_i\dot q_i-L(q,\dot q),\label{legendre}
\end{equation}
we consider the so-called {\it total} or {\it extended} Hamilton function \cite{Dirac:2001}
\begin{align}
H_T(q,p)&=\sum_{j=M+1}^N p_j\dot q_j(p,q)
-G(q,\dot q_{M+1}(p,q),\ldots,\dot q_N(p,q))
+\sum_{i=1}^M\lambda_i \phi_i(q,p)\nonumber\\
&=H(q,p)+\sum_{i=1}^M\lambda_i \phi_i(q,p),
\end{align}
where the $\lambda_i$, $i=1,\ldots M$, are Lagrange multipliers taking care of the
primary constraints and the $\dot q_i(p,q)$ are the solutions to
Eq.~(\ref{kanonimp'}) for $i=M+1,\ldots,N$.

   The constraints $\phi_i$, $i=1,\ldots,M$,  have to be zero throughout all time.
   For consistency, $\dot\phi_i$ must also be zero.
   According to this statement, the time evolution of the primary constraints $\phi_i$ is given by the
Poisson bracket with the total Hamilton function,
leading to the consistency conditions
\begin{equation}
\{\phi_i,H_T\}=\{\phi_i,H\}+\sum_{j=1}^M\lambda_j\{\phi_i,\phi_j\}\approx 0,\quad i=1,\ldots, M.
\end{equation}
   The ``weak'' equality sign refers to the fact that the conditions hold only after the evaluation of Poisson brackets.
   Either all the $\lambda_i$ can be determined from these equations, or
new constraints arise.
   The number of these secondary constraints corresponds to the number
of $\lambda$'s (or linear combinations thereof) which could not
be determined.
   Again one demands the conservation in time of these (new) constraints and
tries to solve the remaining $\lambda$'s from these equations, etc.
   The number of physical degrees of freedom is given by the initial number of
degrees of freedom (coordinates plus momenta) minus the number of constraints.
   In order for a theory to be consistent, the chain of new
constraints has to terminate such that at the end of the procedure
the correct number of degrees of freedom has been generated.
   Using this consistency condition, we could have some restrictions on the possible interactions terms.

\section{Lagrangian}
\label{section_lagrangian}

   The vector mesons are described by eight real vector fields $V^\mu_a$, and the spin-$\frac{1}{2}$ baryons
by eight complex Dirac fields $\Psi_a$ (and adjoint fields $\Psi_a^\dagger$).
   The behavior of the fields under infinitesimal \emph{global} SU(3) transformations is given by
\begin{subequations}
\begin{align}
\label{Vtrans}
V_a^\mu&\mapsto V_a^\mu+f_{abc}\epsilon_b V^\mu_c,\\
\label{Psitrans}
\Psi_a&\mapsto \Psi_a+f_{abc}\epsilon_b \Psi_c,\\
\label{Psidaggertrans}
\Psi^\dagger_a&\mapsto \Psi^\dagger_a+f_{abc}\epsilon_b \Psi^\dagger_c,
\end{align}
\end{subequations}
where $f_{abc}$ denotes the structure constants of SU(3).
   Equations (\ref{Vtrans})--(\ref{Psidaggertrans}) express the fact that
the corresponding fields in each case transform according to the adjoint representation
as SU(3) octets.

   The most general effective Lagrangian for a system of a massive vector-meson octet interacting with a massive baryon octet can
be written as
\begin{equation}
\label{lagrangian}
\mathcal{L}=\mathcal{L}_\text{1}+\mathcal{L}_{\nicefrac{1}{2}} + \mathcal{L}_{\text{int}}+\ldots.
\end{equation}
   Let us first comment on the terms which are not explicitly shown in Eq.~(\ref{lagrangian}).
   The ellipses stand for an infinite string of ``nonrenormalizable'' higher-order interactions as well as for interactions
with other hadrons.
   We make the assumption that the ``nonrenormalizable'' interactions are suppressed by powers of some large scale and concentrate,
at the present time, on the leading-order Lagrangians $\mathcal{L}_\text{1}$, $\mathcal{L}_{\nicefrac{1}{2}}$, and $\mathcal{L}_{\text{int}}$ given by\footnote{For
the sake of simplicity, we suppress subscripts 0 denoting the bare parameters and the bare fields.}
\begin{subequations}
\begin{align}
\label{lagrangian1}
\mathcal{L}_{1}&=-\frac{1}{4} V_{a \mu\nu} V_a^{\mu\nu}
+ \frac{M_{V}^2}{2} V_{a\mu} V_a^{\mu}
- gf_{abc}(\partial_{\mu}V_{a\nu}) V_b^{\mu} V_c^{\nu}- \frac{g^2}{4} f_{abc}f_{ade}V_{b\mu}V_{c\nu}V_d^{\mu} V_e^{\nu},\\
\label{lagrangian2}
\mathcal{L}_{\nicefrac{1}{2}}&=\frac{i}{2}\bar{\Psi}_{a}\gamma^\mu\partial_\mu \Psi_a
-\frac{i}{2}(\partial_\mu\bar{\Psi}_{a})\gamma^\mu\Psi_a-M_B\bar{\Psi}_a\Psi_a,\\
\label{lagrangian3}
\mathcal{L}_\text{int}&=-i \text{G}_\text{F}f_{abc} \bar{\Psi}_{a}\gamma^{\mu}\Psi_{b} V_{c\mu} +\text{G}_\text{D}d_{abc}  \bar{\Psi}_{a}\gamma^{\mu}\Psi_{b}V_{c\mu}.
\end{align}
\end{subequations}
   We have taken these Lagrangians to be invariant under the infinitesimal \emph{global} SU(3) transformations
of Eqs.~(\ref{Vtrans})--(\ref{Psidaggertrans}).
   As a result, the members of the vector-meson octet have a common mass $M_V$, and the mass
of the baryon octet is denoted by $M_B$.
   In Eq.~(\ref{lagrangian1}), the field-strength tensor is defined as
$V_{a\mu\nu}=\partial_\mu V_{a\nu}-\partial_\nu V_{a\mu}$.
   Moreover, for the vector-meson self interaction, the constraint analysis of Refs.~\cite{Neiser:2011,Bijnens:2014fya}
has already been incorporated, leading to a reduction from originally five independent couplings to one single coupling $g$.
   The Lagrangian $\mathcal{L}_{1}$ is hence nothing else but the massive Yang-Mills model.
   Owing to the assumed SU(3) symmetry, the interaction between the vector-meson octet and the baryon octet,
Eq.~(\ref{lagrangian3}), can be parametrized in terms of two couplings $\text{G}_\text{F}$ and $\text{G}_\text{D}$,
where $d_{abc}$ denotes the $d$ symbols of SU(3).
   Note that in SU(2) a structure proportional to $d$ symbols does not exist.
   The interaction Lagrangian of Eq.~(\ref{lagrangian3}) represents the analog of the $D$ and $F$ terms in the interaction
of the Goldstone-boson octet with the baryon octet \cite{Krause:1990xc}.
   To summarize, at the Lagrangian level we start with a massive Yang-Mills Lagrangian for the vector mesons
involving one dimensionless coupling $g$ as justified in Refs.~\cite{Djukanovic:2010tb,Neiser:2011,Bijnens:2014fya}.
   The interaction between the vector-meson octet and the baryon octet contains two SU(3) structures with
couplings $\text{G}_\text{F}$ and $\text{G}_\text{D}$.

\section{Classical constraint analysis}
\label{section_classical_constraint_analysis}
   The Lagrangian description of spin-1 particles in terms of vector fields $V^\mu$ contains too
many degrees of freedom, namely, four instead of three fields.
   In other words, we need constraints to eliminate the redundant degrees of freedom.
   We perform the transition to the Hamiltonian formulation and investigate whether the Lagrangians
of Eqs.~(\ref{lagrangian1})--(\ref{lagrangian3}) lead to a consistent interaction with the correct
number of degrees of freedom.
   This is the case as soon as one has obtained the appropriate number of constraint equations
and simultaneously can solve for all the Lagrange multipliers.
   Moreover, to include the fermionic degrees of freedom at a ``classical'' level, we treat
the fields $\Psi_{\alpha a}$ and $\Psi^\ast_{\alpha a}$ as independent Grassmann fields related by
formal complex conjugation \cite{Zinn-Justin:2005}.
   The indices $\alpha$ and $a$ refer to the Dirac-spinor components and the SU(3)-flavor components,
respectively.
   We will also need the corresponding generalization of the Poisson bracket which is given in
Appendix \ref{appendix_generalized_Poisson-brackets}.

   Before performing the Dirac constraint analysis, let us count the number of DOF in the Hamiltonian
framework, where the fields and the momentum fields are regarded as independent variables.
   Starting from the vector fields $V^\mu_a$ together with the conjugate momentum fields $\pi^\mu_a$,
we have $8\cdot 4+8\cdot 4=64$ fields, whereas for 8 spin-1 fields we only need $8\cdot 3+8\cdot 3=48$
independent fields.
   This means that we need to produce 16 constraints.
   For the spin-$\nicefrac{1}{2}$ fields we start with $8\cdot 4+8\cdot 4=64$ fields $\Psi_{\alpha a}$ and $\Psi_{\alpha a}^\ast$
and 64 conjugate momentum fields $\Pi_{\Psi\alpha a}$ and $\Pi_{\Psi^\ast\alpha a}$.
   Indeed, we expect $8\cdot 2\cdot 2$ (fields) plus $8\cdot 2\cdot 2$ (conjugate momentum fields)
independent DOF.
   In other words, we need to produce 64 constraints.

   In the canonical formalism the momentum field variables conjugate to the field variables are given by
\begin{subequations}
\begin{align}
\label{piamu}
\pi_{a\mu}&=\frac{\partial\mathcal{L}}{\partial\dot{V}_a^\mu}
=\frac{\partial{\mathcal L}_1}{\partial\dot{V}_a^\mu},\\
\label{PiPsia}
\Pi_{\Psi\alpha a}&=\frac{\partial^L{\mathcal L}}{\partial\dot{\Psi}_{\alpha a}}=
\frac{\partial^L{\mathcal L}_{\nicefrac{1}{2}}}{\partial\dot{\Psi}_{\alpha a}}
=-\frac{i}{2}\Psi^\ast_{\alpha a},\\
\label{PiPsidaggera}
\Pi_{\Psi^\ast\alpha a}&=\frac{\partial^L\mathcal{L}}{\partial \dot{\Psi}_{\alpha a}^\ast}
=\frac{\partial^L{\mathcal L}_{\nicefrac{1}{2}}}{\partial \dot{\Psi}_{\alpha a}^\ast}
=-\frac{i}{2}\Psi_{\alpha a}.
\end{align}
\end{subequations}
   Here, we follow the convention of Ref.~\cite{Henneaux:1992} and define both conjugate
momentum fields $\Pi_{\Psi\alpha a}$ and $\Pi_{\Psi^\ast\alpha a}$ in terms of left derivatives.
   As a result, $\Pi_{\Psi^\ast\alpha a}=-\Pi_{\Psi\alpha a}^\ast$.
   Using these relations, we immediately see that the ``velocities'' cannot be expressed in terms of the
``momenta.''
   In this case, we cannot immediately pass over from the Lagrangian description in terms of fields and velocity fields
to the Hamiltonian description in terms of fields and momentum fields.
   To define the Hamiltonian of the system, we introduce $3$ equations for so-called primary constraints \cite{Dirac:2001},
\begin{subequations}
\begin{align}
\label{pc1}
{\theta}_{Va}&=\pi_{a0} + gf_{abc}V_{b0}V_{c0}\approx 0,\\
\label{pc2}
\chi_{1\alpha a}&= \Pi_{\Psi\alpha a}+\frac{i}{2}\Psi^\ast_{\alpha a}\approx0,\\
\label{pc3}
\chi_{2\alpha a}&=\Pi_{\Psi^\ast\alpha a}+\frac{i}{2}\Psi_{\alpha a} \approx0,
\end{align}
\end{subequations}
where $a=1,\ldots, 8$ and $\alpha=0,1,2,3$.
   In these equations, a relation such as ${\theta}_{Va}\approx0$ denotes a weak equation in Dirac's sense \cite{Dirac:2001},
namely that one must not use one of these constraints before working out a Poisson bracket.
   In total, Eqs.~(\ref{pc1})--(\ref{pc3}) result in 8 constraints for the vector mesons and $8\cdot 4+8\cdot 4=64$ constraints
for the baryons.
   We introduce a set of unknown Lagrange multiplier functions, i.e.~$\{\lambda_{1\alpha a}, \lambda_{2\alpha a}, \lambda_{Va}\}$,
for each constraint one Lagrange multiplier, and define a constraint Hamiltonian (density) $\mathcal{H}_\text{c}$
through
\begin{equation}
\mathcal{H}_\text{c}=\lambda_{1\alpha a}\chi_{1\alpha a}+\lambda_{2\alpha a}\chi_{2\alpha a}+\lambda_{V a}\theta_{Va}.
\end{equation}
   We make use of $\chi_{1\alpha a}^\ast=-\chi_{2\alpha a}$ and ${\theta}_{Va}^\ast={\theta}_{Va}$, and require
$\mathcal{H}_\text{c}$ to be real.
   Noting that $\lambda_{1\alpha a}$, $\lambda_{2\alpha a}$, $\chi_{1\alpha a}$, and $\chi_{2\alpha a}$ are all
odd functions (see Appendix \ref{appendix_generalized_Poisson-brackets}), this
implies for the Lagrange multipliers $\lambda_{1\alpha a}^\ast=\lambda_{2\alpha a}$
and $\lambda_{Va}^\ast=\lambda_{Va}$.
   The so-called total or extended Hamiltonian (density) is constructed in terms of a Legendre transformation and the
constraint Hamiltonian (density) $\mathcal{H}_\text{c}$ as
\begin{equation}
\mathcal{H}_{T}=\mathcal{H}_{1} + \mathcal{H}_{\nicefrac{1}{2}} + \mathcal{H}_\text{int}+\mathcal{H}_\text{c}.
\end{equation}
   The explicit expressions for the Hamiltonian densities are given in Appendix \ref{appendix_hamiltonian_densities}.

\begin{center}
\begin{table}[ht]
\caption{\label{opt} Counting the DOF for the free vector, Dirac, and interacting theories, respectively.}
\centering
\begin{tabular}{l c c c}
\hline\hline
Case & Total DOF & Constraints & Physical DOF \\ [.1ex]
Free vector fields & 64 & 16 & 48 \\
Free Dirac fields & 128 & 64 & 64 \\
Interacting theory
 & 192 & 80 & 112 \\ [.1ex]
\hline\hline
\end{tabular}
\label{table:nonlin}
\end{table}
\end{center}

   The requirement that Eqs.~(\ref{pc1})--(\ref{pc3}) have to be zero throughout all time results in
\begin{subequations}
\begin{align}
\label{poisson1}
\left\{\theta_{Va}, H_{T}\right\}&=\partial _{i}\pi^{i}_a+M_{V}^2 V_{a0}-gf_{abc}\pi_b^iV_{ci}+\ldots\nonumber\\
&\equiv\vartheta_{Va}\approx 0, \\
\label{poisson2}
\{\chi_{1\alpha a}, H_T\}&=i (\partial_i\Psi^\ast_{\beta a})(\gamma^0\gamma^i)_{\beta\alpha}+M_B\Psi_{\beta a}^\ast\gamma^0_{\beta\alpha}+\ldots+
i\lambda_{2\alpha a}=0,\\
\label{poisson3}
\{\chi_{2\alpha a}, H_T\}&=i(\gamma^0\gamma^i)_{\alpha\beta}\partial_i\Psi_{\beta a}-M_B\gamma^0_{\alpha\beta}
\Psi_{\beta a}+\ldots+i\lambda_{1\alpha a}=0,
\end{align}
\end{subequations}
where $H_T=\int d^3 x\, {\cal H}_T$ is the total Hamilton function.
    The full expressions for the Poisson brackets are displayed in Appendix \ref{appendix_full_results}.

   From Eqs.~(\ref{poisson2}) and (\ref{poisson3}) we can solve for the Lagrange multipliers
$\lambda_{2\alpha a}$ and $\lambda_{1\alpha a}$, respectively.
   In other words, in the fermionic sector, we have produced the correct number of constraints,
namely 64, and have also determined the 64 Lagrange multipliers, without any conditions for
the coupling constants $\text{G}_\text{F}$ and $\text{G}_\text{D}$.
   Equation (\ref{poisson1}) is a so-called secondary constraint, and, therefore, we obtain 8 additional constraints.
   Also these constraints have to remain conserved with time.
   In fact, evaluating the Poisson bracket of $\vartheta_{Va}$ and $H_T$ results in an equation for the Lagrange multiplier
$\lambda_{Va}$ [see Eq.~(\ref{Poisson_varthetaVa})].
   By inserting the results for the fermionic Lagrange multipliers $\lambda_{1\alpha a}$ and $\lambda_{2\alpha a}$,
at this stage, we have solved for all the Lagrange multipliers and have generated the correct number of constraints.
   The results for the number of DOF are summarized in Table \ref{table:nonlin}.

   As a result of Dirac's constraint analysis, at the classical level we have a self-consistent theory with
the correct number of constraints and thus the correct number of physical DOF without any relation among the couplings.
   In other words, at the classical level, the constants $g$, $\text{G}_\text{F}$, and $\text{G}_\text{D}$ may be regarded
as independent parameters.

\section{Renormalizability}
\label{section_renormalizability}

   We have seen in Sec.~\ref{section_classical_constraint_analysis} that,
at the classical level, the leading-order Lagrangians of Eqs.~(\ref{lagrangian1})--(\ref{lagrangian3})
provide consistent interactions with the correct number of DOF.
   In particular, at this stage, the coupling constants $g$, $\text{G}_\text{F}$, and $\text{G}_\text{D}$ are independent
parameters of the theory.
   When using these Lagrangians in perturbative calculations beyond the tree level, we will
encounter ultraviolet divergences which need to be compensated in the process of renormalization \cite{Collins:1984}.
   At the one-loop level, the perturbative renormalizability condition states that all the divergent parts
of the one-loop diagrams must be canceled by the tree-level diagrams originating from the corresponding counter-term
Lagrangian.
   Since we are working with the {\it most general} effective Lagrangian satisfying the underlying symmetries,
perturbative renormalizability in the sense of EFT requires that the ultraviolet divergences of loop diagrams
can be absorbed in the redefinition of the masses, coupling constants, and fields of the effective Lagrangian
\cite{Weinberg:1978kz,Weinberg:mt}.
   However, it may turn out that this is only possible if certain {\it additional} relations exist among
the coupling constants.

\subsection{Counter-term Lagrangian}
   In order to see whether the couplings $g$, $\text{G}_\text{F}$, and $\text{G}_\text{D}$ are
related through renormalizability, we investigate the vector-meson self energy as well as the $VVV$ and $VVVV$
vertex functions at the one-loop level.
   To identify the counter-term Lagrangian, we relate the bare fields $\Psi_0$ and $V_0^\mu$ to the renormalized
fields $\Psi$ and $V^\mu$,
\begin{equation}
\Psi_0=\sqrt{Z_\Psi}\Psi,\quad V_0^{\mu}=\sqrt{Z_V}V^{\mu},
\end{equation}
and express the bare parameters and the wave-function renormalization constants in terms of
the renormalized parameters,
\begin{subequations}
\begin{align}
g_0&=g+\delta g,\\
\text{G}_\text{F0}&=\text{G}_\text{F}+\delta\text{G}_\text{F},\\
\text{G}_\text{D0}&=\text{G}_\text{D}+\delta\text{G}_\text{D},\\
M_{B0}&=M_B+\delta M_B,\\
M^2_{V0}&=M^2_V+\delta M^2_V,\\
Z_\Psi&=1+\delta Z_\Psi,\\
Z_{V}&=1+\delta Z_{V}.
\end{align}
\end{subequations}
   The functions $\delta g$ etc.~depend on all the renormalized parameters and on the
renormalization condition.
   The counter-term Lagrangian is then given by
\begin{equation}
\label{counter-term_Lagrangian}
\begin{aligned}
\mathcal{L}_\text{ct}&=
-\frac{1}{4}\delta Z_V V_{a\mu\nu} V_a^{\mu\nu}+\frac{1}{2}\delta \{M_V^2\}{V}_{a\mu}V_a^{\mu}\\
&\quad-\frac{1}{4}\delta \{g^2\}f_{abc}f_{ade}V_{b\mu}V_{c\nu} V_d^{\mu}V_e^{\nu}-\delta \{g\}f_{abc}(\partial_{\mu}V_{a\nu}) V_b^{\mu}V_c^{\nu}\\
&\quad+\frac{i}{2}\delta Z_\Psi\left(\bar{\Psi}_a \gamma^{\mu}\partial_{\mu}\Psi_a
-(\partial_\mu\bar{\Psi}_{a})\gamma^\mu\Psi_a\right)
-\delta \{M_B\}\bar{\Psi}_a\Psi_a\\
&\quad-i\delta \{\text{G}_\text{F}\}f_{abc}\bar{\Psi}_{a}\gamma^{\mu}\Psi_{b} V_{c\mu}+\delta \{\text{G}_\text{D}\}d_{abc} \bar{\Psi}_{a}\gamma^{\mu}\Psi_{b}V_{c\mu},
\end{aligned}
\end{equation}
where we display only those terms generated from the Lagrangians in Eqs.~(\ref{lagrangian1})--(\ref{lagrangian3}).
   The expressions for the counter-term functions $\delta \{M_V^2\}$ etc.~are given in Appendix \ref{appendix_counter-term_functions}.

\subsection{Derivation of the conditions}
  We now investigate the divergent parts of all one-loop contributions to the self energies and the vertex functions shown in Fig.~1.\footnote{The
one-loop contributions involving internal vector-meson lines, generate expressions of orders $g^3$ and $g^4$ which need to be canceled
by separate counter-term contributions.}
   Omitting for simplicity both flavor and Lorentz indices, the relation between the unrenormalized (or bare)
and renormalized proper vertex functions involving three and four vector fields, respectively, reads \cite{Itzyckson:1980}
\begin{subequations}
\begin{align}
\label{Gamma3VR}
\Gamma_{3V}^R &= Z_V^{\frac{3}{2}} \, \Gamma_{3V}^0 ,\\
\label{Gamma4VR}
\Gamma_{4V}^R &= Z_V^{2} \, \Gamma_{4V}^0 ,
\end{align}
\end{subequations}
where $\Gamma_{3V}^0$ and $\Gamma_{4V}^0$ are unrenormalized vertex functions and $Z_V$ is the
wave-function renormalization constant of the vector field.
   The vertex functions and the wave-function renormalization constant may be expanded
in powers of $\hbar$,
\begin{subequations}
\begin{align}
\label{Gamma0expanded}
\Gamma^0 & = \Gamma^{\text{tree}}+\hbar \,\Gamma^{1\,\text{loop}}+{\cal O}(\hbar^2),\\
\label{ZVexpanded}
Z_V &= 1+\hbar\, \delta Z_V^{1\,\text{loop}}  +{\cal O}(\hbar^2).
\end{align}
\end{subequations}
   Substituting Eqs.~(\ref{Gamma0expanded}) and (\ref{ZVexpanded}) into Eqs.~(\ref{Gamma3VR})
and (\ref{Gamma4VR}), we obtain the expansions
\begin{subequations}
\begin{align}
\label{Gamma3VRexpanded}
\Gamma_{3V}^R &=\Gamma_{3V}^{\text{tree}}+\hbar\left(\Gamma_{3V}^{1\,\text{loop}}+\frac{3}{2}\delta Z_V^{1\,\text{loop}}\,\Gamma_{3V}^{\text{tree}}\right)+{\cal O}(\hbar^2),\\
\label{Gamma4VRexpanded}
\Gamma_{4V}^R &=\Gamma_{4V}^{\text{tree}}+\hbar\left(\Gamma_{4V}^{1\,\text{loop}}+2 \delta Z_V^{1\,\text{loop}}\,\Gamma_{4V}^{\text{tree}}\right)+{\cal O}(\hbar^2).
\end{align}
\end{subequations}
   The tree-level diagrams have the following form,
\begin{subequations}
\begin{align}
\label{Gamma3Vtree}
\Gamma_{3V}^{\text{tree}}&= g_0 S_{3V},\\
\label{Gamma4Vtree}
\Gamma_{4V}^{\text{tree}}&= g_0^2 S_{4V},
\end{align}
\end{subequations}
where $S_{3V}$ and $S_{4V}$ denote both Lorentz and flavor structures.
   The corresponding divergent parts of the loop diagrams in Fig.~1 contain the same Lorentz structures.
   In terms of the renormalized coupling $g$, the bare coupling $g_0$ can be written as
\begin{equation}
\label{g0expanded}
g_0=g+\hbar\delta g^{1\,\text{loop}} +{\cal O}(\hbar^2),
\end{equation}
where $\delta g^{1\,\text{loop}}$ is the one-loop counter term.
   Using Eq.~(\ref{g0expanded}) in Eqs.~(\ref{Gamma3Vtree}) and (\ref{Gamma4Vtree}), we obtain from Eqs.~(\ref{Gamma3VRexpanded}) and (\ref{Gamma4VRexpanded})
the expressions
\begin{subequations}
\begin{align}
\label{Gamma3VRresult}
\Gamma_{3V}^R &=g S_{3V}+\hbar\delta g^{1\,\text{loop}}\, S_{3V}
+\hbar \left(\Gamma_{3V}^{1\,\text{loop}}+\frac{3}{2}\delta Z_V^{1\,\text{loop}} g S_{3V}\right)+{\cal O}(\hbar^2),\\
\label{Gamma4VRresult}
\Gamma_{4V}^R &= g^2 S_{4V}+ 2 \hbar g \delta g^{1\,\text{loop}}\, S_{4V}+\hbar\left(\Gamma_{4V}^{1\,\text{loop}}+2 \delta Z_V^{1\,\text{loop}} g^2 S_{4V}\right)+{\cal O}(\hbar^2).
\end{align}
\end{subequations}
   The left-hand sides of Eqs.~(\ref{Gamma3VRresult}) and (\ref{Gamma4VRresult}), i.e., $\Gamma_{3V}^R$ and
$\Gamma_{4V}^R$, are finite.
   On the right-hand sides, the tree contributions, i.e., $g S_{3V}$ and $g^2 S_{4V}$, are also finite.
   In Eq.~(\ref{Gamma3VRresult}), $\delta g^{1\,\text{loop}}$ must cancel the divergent parts of the expression inside the parentheses,
which depend on the coupling constants.
   Otherwise, the theory would not be renormalizable (in the sense of effective field theory).
   On the other hand, for the same reason the same $\delta g^{1\,\text{loop}}$ has to cancel the divergences inside the parentheses of Eq.~(\ref{Gamma4VRresult}),
which also depend on the coupling constants, but with a different functional form.
   These two conditions for $\delta g^{1\,\text{loop}}$ ultimately lead to relations among the coupling constants.
   From Eqs.~(\ref{Gamma3VRresult}) and (\ref{Gamma4VRresult}) we obtain for the terms linear in $\hbar$ the conditions
\begin{subequations}
\begin{align}
\label{cond1}
\delta g^{1\,\text{loop}}\, S_{3V} +\left(\Gamma_{3V}^{1\,\text{loop}}+\frac{3}{2} \delta Z_V^{1\,\text{loop}} g S_{3V}\right)&=0,\\
\label{cond2}
2 g \delta g^{1\,\text{loop}}\, S_{4V}+\left(\Gamma_{4V}^{1\,\text{loop}}+2 \delta Z_V^{1\,\text{loop}} g^2 S_{4V}\right)&=0.
\end{align}
\end{subequations}

\subsection{SU(2)}
   Before addressing the universality principle in SU(3), we first want to reproduce the case of SU(2) \cite{Djukanovic:2004mm}.
   To that end, we consider the diagrams of Fig.~1.
   Introducing Lorentz- and isospin indices, the vector-meson self energy may be parameterized as \cite{Djukanovic:2009zn}
\begin{equation}
\Pi^{\mu\nu}_{ij}(p)=\delta_{ij}\left[g^{\mu\nu}\Pi_1(p^2)+p^\mu p^\nu\Pi_2(p^2)\right].
\end{equation}
   Using dimensional regularization, the result for the divergent part of the self-energy diagram reads
\begin{subequations}
\begin{align}
\label{selfPi1}
\Pi_1^{\text{div}}(p^2)&=-\frac{\lambda}{12\pi^2}g_{VNN}^2 p^2,\\
\label{selfPi2}
\Pi_2^{\text{div}}(p^2)&=\frac{\lambda}{12\pi^2}g_{VNN}^2,
\end{align}
\end{subequations}
where $\lambda$ is given by
\begin{equation}
\lambda=\frac{1}{16\pi^2}\left\{\frac{1}{D-4}-\frac{1}{2}\left[\text{ln}(4\pi)+\Gamma'(1)+1\right]\right\},
\end{equation}
with $D$ the number of spacetime dimensions.
   The wave-function renormalization constant is related to the residue of the propagator at the pole,
$p^2=M_V^2$.
   In terms of the self-energy function $\Pi_1(p^2)$ it is
given by
\begin{eqnarray}
Z_{V}=\frac{1}{1-\Pi'_1(M^2_V)}.
\label{cwf}
\end{eqnarray}
   Since we are working at one-loop order, $Z_V$ can be written as
\begin{eqnarray}
Z_V = 1+\Pi'_1(M_V^2)+{\cal O}(\hbar^2),
\label{resi}
\end{eqnarray}
where ${\cal O}(\hbar^2)$ stands for two-loop corrections.
   Inserting Eq.~(\ref{selfPi1}) into Eq.~(\ref{resi}), we have for the
part proportional to $\lambda$,
\begin{eqnarray}
\delta Z_V^\lambda=-\frac{\lambda}{12\pi^2}g_{VNN}^2.
\label{wf}
\end{eqnarray}
   The divergent parts of the one-loop contributions to the three- and four-vector vertex functions read, respectively,
\begin{subequations}
\begin{align}
\label{3vf}
\Gamma_{ijk}^{\mu \nu \rho\,\text{div}}(p_1,p_2,p_3)
&=\epsilon_{ijk}\frac{\lambda}{12\pi^2} g_{VNN}^3 [g^{\mu \nu}(p_1-p_2)^\rho +g^{\mu \rho}(p_2-p_3)^\nu+g^{\nu \rho}(p_3-p_1)^\mu], \\
\label{4vf}
\Gamma_{ijkl}^{\mu \nu \rho \sigma\,\text{div}}(p_1,p_2,p_3,p_4)
&=-\frac{i\lambda}{12\pi^2}g_{VNN}^4[(2\delta_{ij}\delta_{kl}-\delta_{ik}\delta_{jl}-\delta_{jk}\delta_{il})g^{\mu \nu}g^{\rho \sigma}
\nonumber\\
 &\quad +(2\delta_{ik}\delta_{jl}-\delta_{il}\delta_{jk}-\delta_{ij}\delta_{kl})g^{\mu \rho}g^{\nu \sigma}\nonumber\\
 &\quad +(2\delta_{il}\delta_{jk}-\delta_{ik}\delta_{jl}-\delta_{ij}\delta_{kl})g^{\mu \sigma}g^{\nu \rho}].
\end{align}
\end{subequations}
   Substituting Eq.~(\ref{wf}) and Eqs.~(\ref{3vf}) and (\ref{4vf}) in Eqs.~(\ref{cond1}) and (\ref{cond2}),
we obtain the following two expressions for $\delta g^\lambda$,
\begin{subequations}
\begin{align}
\label{deltag1}
\delta g^\lambda&=\frac{\lambda}{8\pi^2}g g_{VNN}^2-\frac{\lambda}{12\pi^2}g_{VNN}^3,\\
\label{deltag2}
\delta g^\lambda&=\frac{\lambda}{12\pi^2}g g_{VNN}^2-\frac{\lambda}{24\pi^2}\frac{g_{VNN}^4}{g}.
\end{align}
\end{subequations}
   In a self-consistent theory the two expressions for $\delta g^\lambda$ must coincide.
   This is true for the trivial case $g_{VNN}=0$, i.e., for a theory without
lowest-order interaction between the vector mesons and the nucleon.
   The non-trivial solution is given by
\begin{eqnarray}
g_{VNN}=g,
\label{solution}
\end{eqnarray}
which corresponds to the universality principle in SU(2).
   Consequently, from the EFT perspective, the universal coupling $g_{VNN}=g$ is a result of the consistency conditions
imposed by the requirement of perturbative renormalizability (see also Ref.\ \cite{Djukanovic:2004mm}).
\begin{figure}
\includegraphics[width=\textwidth]{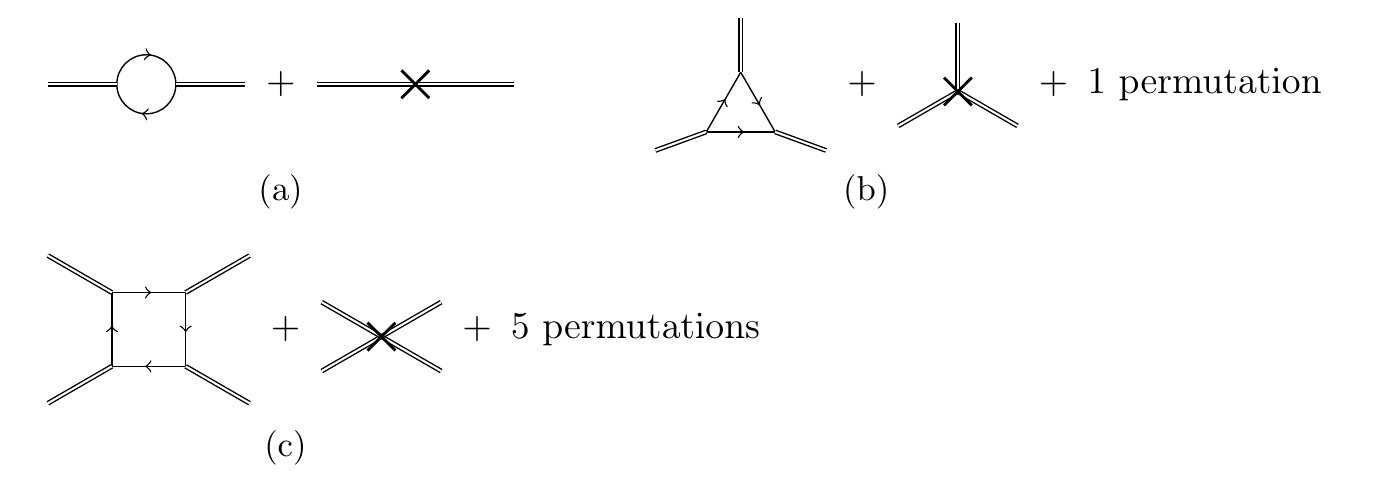}
\caption{(a) Nucleon-loop contribution to the vector-meson self-energy diagram,
(b) one-loop contributions to the three-vector vertex function, and
(c) four-vector vertex function. Single and double lines correspond to fermions and bosons, respectively.}
\end{figure}

\subsection{SU(3)}
   In this section, we look for relations among the renormalized coupling constants $\text{G}_\text{F}$, $\text{G}_\text{D}$, and $g$
of the SU(3) Lagrangian of Eq.~(\ref{lagrangian}).
   The method is similar to the case of SU(2), but this time we need to consider different
SU(3) flavor combinations in order to disentangle the conditions for $\text{G}_\text{F}$ and $\text{G}_\text{D}$.

   The divergent part of the self-energy diagram in Fig.~1 is given by
\begin{equation}
\Pi_{ab}^{\mu \nu\,\text{div}}(p)=-\frac{\lambda}{6\pi^2}(5\text{G}_\text{D}^2+9\text{G}_\text{F}^2)\delta_{ab}(g^{\mu \nu}p^2-p^\mu p^\nu),
\label{self2}
\end{equation}
from which we obtain by using Eq.~(\ref{resi})
\begin{equation}
\delta Z_{V}^\lambda=-\frac{\lambda}{6\pi^2}(5\text{G}_\text{D}^2+9\text{G}_\text{F}^2).
\label{cwf2}
\end{equation}
   In contrast to the SU(2) case, we will calculate the four-vector vertex function for
two different combinations of flavor indices to obtain two expressions for $\delta g^\lambda$.
   To be specific, we consider the combinations $(a,b,c,d)=(1,3,1,3)$ and $(a,b,c,d)=(1,6,1,6)$, respectively, and obtain
\begin{subequations}
\begin{align}
\label{4vfSU3_1}
\Gamma^{\mu \nu \rho \sigma\,\text{div}}_{1313}&=
\frac{i\lambda}{54\pi^2}(11\text{G}_\text{D}^4+90\text{G}_\text{D}^2\text{G}_\text{F}^2+27\text{G}_\text{F}^4)(g^{\mu \rho}g^{\nu \sigma}
+g^{\rho \sigma}g^{\mu \nu}-2g^{\nu \rho}g^{\mu \sigma}),\\
\label{4vfSU3_2}
\Gamma^{\mu \nu \rho \sigma\,\text{div}}_{1616}&=
\frac{i\lambda}{216\pi^2}(35\text{G}_\text{D}^4+90\text{G}_\text{D}^2\text{G}_\text{F}^2+27\text{G}_\text{F}^4)(g^{\mu \rho}g^{\nu \sigma}
+g^{\rho \sigma}g^{\mu \nu}-2g^{\nu \rho}g^{\mu \sigma}).
\end{align}
\end{subequations}
   We now consider Eq.~(\ref{cond2}) for the combinations $(a,b,c,d)=(1,3,1,3)$ and $(a,b,c,d)=(1,6,1,6)$, respectively,
and make use of the results of Eq.~(\ref{cwf2}) and Eqs.~(\ref{4vfSU3_1}) and (\ref{4vfSU3_2}) as well as the
tree-level Feynman rule of Table \ref{table:Fd}.
   We obtain two results for $\delta g^\lambda$,
namely,
\begin{subequations}
\begin{align}
\label{dg1}
\delta g^\lambda&=-\frac{\lambda}{108\pi^2g}(11\text{G}_\text{D}^4+90\text{G}_\text{D}^2\text{G}_\text{F}^2
+27\text{G}_\text{F}^4)+\frac{\lambda}{6\pi^2}(5\text{G}_\text{D}^2+9\text{G}_\text{F}^2)g,\\
\label{dg2}
\delta g^\lambda&=-\frac{\lambda}{108\pi^2g}(35\text{G}_\text{D}^4+90\text{G}_\text{D}^2\text{G}_\text{F}^2
+27\text{G}_\text{F}^4)+\frac{\lambda}{6\pi^2}(5\text{G}_\text{D}^2+9\text{G}_\text{F}^2)g.
\end{align}
\end{subequations}
    In a self-consistent theory, the two expressions for $\delta g^\lambda$ must be equal.
   This implies
\begin{equation}
\text{G}_\text{D}=0,
\label{final}
\end{equation}
because otherwise the interacting theory would not be renormalizable in the perturbative
sense of effective field theory.
   Using $\text{G}_\text{D}=0$, the universality $\text{G}_\text{F}=g$ is obtained in
analogy to the SU(2) case by comparing the expressions for $\delta g^\lambda$ obtained
from the three-vector and four-vector vertex functions, respectively.
   We obtain for the three-vector vertex function
\begin{equation}
\delta g^\lambda=\frac{3\lambda}{4\pi^2}g \text{G}_\text{F}^2 +\frac{\lambda}{2\pi^2} \text{G}_\text{F}^3,
\end{equation}
which needs to be compared with
\begin{equation}
\delta g^\lambda=-\frac{\lambda}{4\pi^2 g}\text{G}_\text{F}^4+\frac{3\lambda}{2\pi^2}\text{G}_\text{F}^2 g
\end{equation}
from the four-vector vertex function.
   As solutions we either obtain $\text{G}_\text{F}=0$ or $\text{G}_\text{F}=g$.
   Consequently, our final result is
\begin{eqnarray}
\text{G}_\text{D}=0 \quad\text{and}\quad \text{G}_\text{F}=\text{g}.
\label{final}
\end{eqnarray}
   The renormalizability analysis thus generates relations among the dimensionless
coupling constants of the most general Lagrangian with a global SU(3) symmetry.
   We end up with a universality principle in SU(3), in which the leading-order Lagrangian
is that of a {\it massive} Yang-Mills theory with a universal coupling $g$.
   Of course, Lagrangians of such type were often used in phenomenological
applications (see, e.g., Refs.\ \cite{Klingl:1997kf,Palomar:2002hk,Oset:2009sw,Khemchandani:2011mf}).
   Indeed, by using the requirement of renormalizability in the sense of EFT, the present analysis
provides a further motivation for the universal coupling of vector mesons as originally
discussed in Ref.~\cite{Sakurai:1960ju} for isospin, baryon number, and
hypercharge (see also Ref.~\cite{Sakurai:1969}).

\section{Conclusions}
\label{section_summary}
   At the classical level, the lowest-order SU(3)-invariant Lagrangians of
Eqs.~(\ref{lagrangian1})--(\ref{lagrangian3}), involving three independent coupling
constants $g$, $\text{G}_\text{F}$, and $\text{G}_\text{D}$, define a self-consistent starting
point for the self interaction of the vector-meson octet as well as the
interaction of the vector-meson octet with the baryon octet.
   This was explicitly shown by using Dirac's method.

   However, the requirement of renormalizability in the sense of effective field theory
implies additional constraints among the renormalized couplings.
   By comparing the expression for $\delta g$ obtained from the $VVV$ vertex function on the one hand
and the $VVVV$-vertex function on the other hand, we were able to show that there are relations among the
renormalized coupling constants $g$, $\text{G}_\text{F}$, and $\text{G}_\text{D}$.
   We found a universal interaction with $g=\text{G}_\text{F}$ and $\text{G}_\text{D}=0$.
   In other words, starting from the most general leading-order Lagrangian invariant under a global
SU(3) transformation, we have seen that, after obtaining a universal coupling, the interaction Lagrangian is that of
a (massive) SU(3) Yang-Mills theory.

\section{Acknowledgments}
   When calculating the loop diagrams, we made use of the packages
FeynCalc \cite{Mertig:1990an} and LoopTools \cite{Hahn:1998yk}.
   This work was supported by the Deutsche Forschungsgemeinschaft (SFB 443).
   The work of Y.~\"U was supported by the Scientific and Technological Research Council of Turkey (T\"UB\.{I}TAK).
   The authors would like to thank D.~Djukanovic and A.~Neiser for useful discussions and comments during the work.
   The authors are also greatly indebted to J.~Gegelia for his very valuable contributions to the work.

\newpage
\begin{appendix}

\section{Generalized Poisson brackets}
\label{appendix_generalized_Poisson-brackets}
   With the inclusion of Grassmann fields, we need a generalization of the standard Poisson brackets.
   Here, we only collect the results needed for the present purposes and
refer the reader to chapter 6 of Ref.~\cite{Henneaux:1992} for more details.
   Let $F$ denote a function of the dynamical variables $\Psi_{\alpha a}$, $\Pi_{\Psi\alpha a}$,
$\Psi^\ast_{\alpha a}$, $\Pi_{\Psi^\ast\alpha a}$, $V_{a\mu}$, and $\pi_{a\mu}$.
   The Grassmann parity $\epsilon_F$ is defined to be equal to 0 if the function consists of monomials of Grassmann variables
of even degree, and the function is then said to be even.
   An odd function has Grassmann parity $\epsilon_F=1$ and consists of monomials of Grassmann variables
of odd degree.
   Any function $F$ can be decomposed into its even and odd components, respectively, $F=F_E+F_O$.
   The Poisson bracket of two functionals (or functions) is defined as
\begin{align}
\{F,G\}&=\int d^3x \left(\frac{\delta F}{\delta V_{a\mu}(\vec {x})}\frac{\delta G}{\delta \pi^\mu_a(\vec x)}
-\frac{\delta F}{\delta \pi^\mu_a(\vec x)}\frac{\delta G}{\delta V_{a\mu}(\vec{x})}\right)\nonumber\\
&\quad+(-)^{\epsilon_F}\int d^3x \left(
\frac{\delta^L F}{\delta \Psi_{\alpha a}(\vec {x})}\frac{\delta^L G}{\delta \Pi_{\Psi\alpha a}(\vec x)}
+\frac{\delta^L F}{\delta \Pi_{\Psi\alpha a}(\vec x)}\frac{\delta^L G}{\delta \Psi_{\alpha a}(\vec {x})}\right.\nonumber\\
&\quad\left.+\frac{\delta^L F}{\delta \Psi_{\alpha a}^\ast(\vec {x})}\frac{\delta^L G}{\delta \Pi_{\Psi^\ast\alpha a}(\vec x)}
+\frac{\delta^L F}{\delta \Pi_{\Psi^\ast\alpha a}(\vec x)}\frac{\delta^L G}{\delta \Psi_{\alpha a}^\ast(\vec {x})}
\right),
\end{align}
where a summation over repeated indices is implied, and $F$ is assumed to have a definite
Grassmann parity $\epsilon_F$.
   We suppress time as an argument of the fields, as they are to be evaluated at the same time.
   The symbol $L$ in the functional derivative indicates that the relevant function entering the
functional has to be moved to the left with an appropriate sign factor resulting from the
necessary permutations.
   The fundamental Poisson brackets are given by
\begin{subequations}
\begin{align}
\{V_{a\mu}(\vec{x}),\pi_{b\nu}(\vec y)\}&=\delta_{ab}\delta_{\mu\nu}\delta^3(\vec x-\vec y),\\
\{\Psi_{\alpha a}(\vec x),\Pi_{\Psi\beta b}(\vec y)\}&=-\delta_{\alpha\beta}\delta_{ab}\delta^3(\vec x-\vec y),\\
\{\Psi_{\alpha a}^\ast(\vec x),\Pi_{\Psi^\ast\beta b}(\vec y)\}&=-\delta_{\alpha\beta}\delta_{ab}\delta^3(\vec x-\vec y).
\end{align}
\end{subequations}
   In addition, the following properties are useful:
\begin{align}
\{F,G\}&=-(-)^{\epsilon_F\epsilon_G}\{G,F\},\\
\{F,GH\}&=\{F,G\}H+(-)^{\epsilon_F\epsilon_G}G\{F,H\},\\
\{F,G\}^\ast&=-\{G^\ast,F^\ast\}.
\end{align}

\section{Hamiltonian densities}
\label{appendix_hamiltonian_densities}
The Hamiltonian densities relevant for the evaluation of the Poisson brackets read
\begin{align*}
{\cal H}_{\frac{1}{2}}&=-\frac{i}{2}\left[\bar{\Psi}_a\gamma^i\partial_i\Psi_a-(\partial_i\bar{\Psi}_a)\gamma^i\Psi_a\right]
+M_B\bar{\Psi}_a\Psi_a,\\
{\cal H}_1&=-\frac{1}{2}\pi_{ai}\pi_a^i+(\partial_i V_{a0})\pi^i_a+\frac{1}{4}V_{aij}V^{ij}_a-\frac{M_V^2}{2}V_{a\mu}V^\mu_a\\
&\quad-gf_{abc}\pi^i_aV_{b0}V_{ci}
+gf_{abc}(\partial^i V_a^j)V_{bi} V_{cj}
+\frac{g^2}{4} f_{abc}f_{ade}V_{bi}V_{cj}V_d^{i} V_e^{j},\\
\mathcal{H}_\text{int}&=i \text{G}_\text{F}f_{abc} \bar{\Psi}_{a}\gamma^{\mu}\Psi_{b} V_{c\mu}
-\text{G}_\text{D}d_{abc}  \bar{\Psi}_{a}\gamma^{\mu}\Psi_{b}V_{c\mu}.
\end{align*}

\section{Results of the constraint analysis}
\label{appendix_full_results}
\begin{align}
\label{Poisson_thetaVa}
\left\{\theta_{Va}, H_{T}\right\}
&=\partial_i\pi^i_a+M_V^2 V_{a0}
-gf_{abc}\pi_b^iV_{ci}
+(-i\text{G}_\text{F}f_{abc}+\text{G}_\text{D}d_{abc})\Psi_b^\dagger\Psi_{c}\nonumber\\
&\equiv\vartheta_{Va}\approx 0,\\
\label{Poisson_varthetaVa}
\left\{\vartheta_{Va}, H_{T}\right\}
&=-M_V^2\partial_iV_a^i+g f_{abc} V_{b0} \partial_i\pi^i_c\nonumber\\
&\quad+g^2f_{abe}f_{ecd}V_{c0}(V_{bi}\pi^i_d-\pi^i_bV_{di})\nonumber\\
&\quad-g^2f_{abe}f_{ecd}V_{bi}[(\partial^j V_{cj}) V_d^i+(\partial^i V^j_c) V_{dj}]
+g^2 f_{abe}f_{ecd}\partial_i(V_c^iV_{dj}V_b^j)\nonumber\\
&\quad-\frac{g^3}{2}f_{abc}f_{dbe}f_{dfg}V_{ci}V_{ej}V_f^i V_g^j\nonumber\\
&\quad+gf_{abc}(i\text{G}_\text{F}f_{cde}-\text{G}_\text{D}d_{cde})V_{bi}\bar{\Psi}_d\gamma^i\Psi_e\nonumber\\
&\quad+(i\text{G}_\text{F}f_{abc}-\text{G}_\text{D}d_{abc})\partial_i(\bar{\Psi}_b\gamma^i\Psi_c)\nonumber\\
&\quad+(-i\text{G}_\text{F}f_{abc}-\text{G}_\text{D}d_{abc})\lambda_{1\alpha b}\Psi^\ast_{\alpha c}
+(i\text{G}_\text{F}f_{abc}-\text{G}_\text{D}d_{abc})\lambda_{2\alpha b}\Psi_{\alpha c}
+M_V^2\lambda_{Va},\\
\label{Poisson_chi1_HT}
\{\chi_{1\alpha a}, H_T\}
&=i(\partial_i\Psi_{\beta a}^{\ast})(\gamma^0\gamma^i)_{\beta\alpha}
+M_B\Psi_{\beta a}^\ast\gamma^0_{\beta\alpha}
-(i\text{G}_\text{F}f_{abc}+\text{G}_\text{D}d_{abc})\Psi^\ast_{\beta b}(\gamma^0\gamma^\mu)_{\beta\alpha} V_{c\mu}
+i\lambda_{2\alpha a}\nonumber\\
&\approx 0,\\
\label{Poisson_chi2_HT}
\{\chi_{2\alpha a}, H_T\}
&=i(\gamma^0\gamma^i)_{\alpha\beta}\partial_i\Psi_{\beta a}
-M_B\gamma^0_{\alpha\beta}\Psi_{\beta a}
+(-i\text{G}_\text{F}f_{abc}+\text{G}_\text{D}d_{abc})(\gamma^0\gamma^\mu)_{\alpha\beta}\Psi_{\beta b} V_{cu}
+i\lambda_{1\alpha a}\nonumber\\
&\approx 0.
\end{align}

\section{Counter-term functions}
\label{appendix_counter-term_functions}
   The expressions for the counter-term functions in Eq.~(\ref{counter-term_Lagrangian}) are given by
\begin{align*}
\delta\{M_B\}&=\delta M_B +\delta Z_\Psi M_B,\\
\delta\{M^2_V\}&=\delta M^2_V+\delta Z_V M_V^2,\\
\delta\{g\}&=\delta g+\frac{3}{2}\delta Z_V g,\\
\delta\{g^2\}&=2\delta g g+2\delta Z_V^2 g^2,\\
\delta\{\text{G}_\text{F}\}&=\delta\text{G}_\text{F}+\left(\delta Z_\Psi+\frac{1}{2}\delta Z_V\right)\text{G}_\text{F},\\
\delta\{\text{G}_\text{D}\}&=\delta\text{G}_\text{D}+\left(\delta Z_\Psi+\frac{1}{2}\delta Z_V\right)\text{G}_\text{D}.
\end{align*}
   We only displayed the terms relevant at leading order in an expansion in $\hbar$, i.e.,
we omitted terms of the type $\delta M_V^2\delta Z_V$ etc.

\pagebreak
\section{Feynman rules}
\label{appendix_feynman_rules}
\begin{center}
\begin{table}[ht]
\caption{Propagators and vertices of Feynman diagrams in SU(2) and SU(3).
Single and double lines correspond to fermions and bosons, respectively.
$a, b, c, d$ correspond to SU(3) octet indices, $i, j, k, l$ and $r, s$
correspond isospin triplet and doublet indices, respectively.}
\centering
\begin{tabular}{l c c c}
\hline\hline
& SU(2) & SU(3) \\ [0ex]
\hline \\[0.3ex]
&
  \pbox{\textwidth}{%
  \begin{tikzpicture}[scale=1]
  \draw [double] (-1,0) -- (1,0);
  \node [below=3pt]{$k$};
  \draw [fill] (-1,0) circle (.05)node [anchor=east]{$\mu,i$};
  \draw [fill] (1,0) circle (.05)node [anchor=west]{$\nu,j$};
  \end{tikzpicture}%
}
&
  \pbox{\textwidth}{%
  \begin{tikzpicture}[scale=1]
  \draw [double] (-1,0) -- (1,0);
  \node [below=3pt]{$k$};
  \draw [fill] (-1,0) circle (.05)node [anchor=east]{$\mu,a$};
  \draw [fill] (1,0) circle (.05)node [anchor=west]{$\nu,b$};
  \end{tikzpicture}%
} \\  [4ex]
Propagators &
$\quad\quad -i \frac{g^{\mu \nu}-\frac{k^{\mu}k^{\nu}}{M^2_V}}{k^2-M^2_V+i\epsilon}\delta_{ij}$
& $\quad\quad -i \frac{g^{\mu \nu}-\frac{k^{\mu}k^{\nu}}{M^2_V}}{k^2-M^2_V+i\epsilon}\delta_{ab}$  \\ [4ex]
&
  \pbox{\textwidth}{%
  \begin{tikzpicture}[scale=1]
  \draw [->-=.55] (-1,0) -- (1,0);
  \node [below=3pt]{$p$};
  \draw [fill] (-1,0) circle (.05)node [anchor=east]{$r$};
  \draw [fill] (1,0) circle (.05)node [anchor=west]{$s$};
  \end{tikzpicture}%
}
&
  \pbox{\textwidth}{%
  \begin{tikzpicture}[scale=1]
  \draw [->-=.55] (-1,0) -- (1,0);
  \node [below=3pt]{$p$};
  \draw [fill] (-1,0) circle (.05)node [anchor=east]{$a$};
  \draw [fill] (1,0) circle (.05)node [anchor=west]{$b$};
  \end{tikzpicture}%
}
\\  [4ex]
&  $\quad\quad  \frac{i}{\slashed{p}-m+i\epsilon}\delta_{rs}$ & $\quad\quad \frac{i}{\slashed{p}-M_B+i\epsilon}\delta_{ab}$  \\ [4ex]
\hline \\ [.3ex]
&
    \pbox{\textwidth}{%
    \begin{tikzpicture}[scale=1]
    \draw [->-=.55] (-1,0) -- (0,0);
    \node [below left] at(-1,.2) {$a$};
    \draw [-<-=.55] (1,0) -- (0,0);
    \node [below right] at(1,.28) {$b$};
    \draw [double] (0,0) -- +(90:.7);
    \node [above right] at(.1,.08) {$\mu,i$};
    \draw [fill] (0,0) circle (.05);
    \end{tikzpicture}%
}
&
    \pbox{\textwidth}{%
    \begin{tikzpicture}[scale=1]
    \draw [->-=.55] (-1,0) -- (0,0);
    \node [below left] at(-1,.2) {$a$};
    \draw [-<-=.55] (1,0) -- (0,0);
    \node [below right] at(1,.28) {$b$};
    \draw [double] (0,0) -- +(90:.7);
    \node [above right] at(.1,.08) {$\mu,c$};
    \draw [fill] (0,0) circle (.05);
    \end{tikzpicture}%
}
\\ [4ex]
&  $\quad\quad  ig \gamma^{\mu}\frac{(\tau_i)_{ba}}{2}$ & $\quad\quad i\gamma^\mu[G_Dd_{abc}+iG_Ff_{abc}]$  \\ [.1ex]
\\ [.3ex]
&
    \pbox{\textwidth}{%
    \begin{tikzpicture}[scale=1]
    \draw [double,-<-=.55] (0,0) -- +(210:.7);
    \node [below left] at(-.6,.08) {$\nu,j,p_2$};
    \draw [double,-<-=.55] (0,0) -- +(-30:.7);
    \node [below right] at(.6,.08) {$\rho,k,p_3$};
    \draw [double,-<-=.55] (0,0) -- +(90:.7);
    \node [above right] at(.1,.08) {$\mu,i,p_1$};
    \end{tikzpicture}%
}
&
    \pbox{\textwidth}{%
    \begin{tikzpicture}[scale=1]
    \draw [double,-<-=.55] (0,0) -- +(210:.7);
    \node [below left] at(-.6,.08) {$\nu,b,p_2$};
    \draw [double,-<-=.55] (0,0) -- +(-30:.7);
    \node [below right] at(.6,-.02) {$\rho,c,p_3$};
    \draw [double,-<-=.55] (0,0) -- +(90:.7);
    \node [above right] at(.1,.08) {$\mu,a,p_1$};
    \end{tikzpicture}%
}
\\ [4ex]
Vertices&  $\begin{array} {lcl}
 & g\epsilon_{ijk}[g^{\mu \nu}(p_1-p_2)^\rho\\
 & +g^{\mu \rho}(p_3-p_1)^\nu+g^{\nu \rho}(p_2-p_3)^\mu]
\end{array}$ & $\begin{array} {lcl}
 & gf_{abc}[g^{\mu \nu}(p_1-p_2)^\rho\\
 & +g^{\mu \rho}(p_3-p_1)^\nu+g^{\nu \rho}(p_2-p_3)^\mu]
\end{array}$  \\ [.1ex]
\\ [.3ex]
&
    \pbox{\textwidth}{%
    \begin{tikzpicture}[scale=1]
    \draw [double] (0,0) -- +(210:.7);
    \node [above left] at(-.6,.08) {$\mu,i$};
    \draw [double] (0,0) -- +(-30:.7);
    \node [above right] at(.6,.08) {$\nu,j$};
    \draw [double] (0,0) -- +(-210:.7);
    \node [below left] at(-.6,-.18) {$\rho,k$};
    \draw [double] (0,0) -- +(30:.7);
    \node [below right] at(.6,-.10) {$\sigma,l$};
    \end{tikzpicture}%
}
&
    \pbox{\textwidth}{%
    \begin{tikzpicture}[scale=1]
    \draw [double] (0,0) -- +(210:.7);
     \node [above left] at(-.6,.08) {$\mu,a$};
    \draw [double] (0,0) -- +(-30:.7);
     \node [above right] at(.6,.08) {$\nu,b$};
    \draw [double] (0,0) -- +(-210:.7);
    \node [below left] at(-.6,-.18) {$\rho,c$};
    \draw [double] (0,0) -- +(30:.7);
     \node [below right] at(.6,-.10) {$\sigma,d$};
    \end{tikzpicture}%
}
\\ [4ex]
&  $\begin{array} {lcl}
 & -ig^2[g^{\mu \nu}g^{\rho \sigma}(2\delta_{ij}\delta_{kl}-\delta_{ik}\delta_{jl}-\delta_{jk}\delta_{il})\\
 & +g^{\mu \rho}g^{\nu \sigma}(2\delta_{ik}\delta_{jl}-\delta_{il}\delta_{jk}-\delta_{ij}\delta_{kl})\\
 & +g^{\mu \sigma}g^{\nu \rho}(2\delta_{il}\delta_{jk}-\delta_{ik}\delta_{jl}-\delta_{ij}\delta_{kl})]\end{array}$ & $\begin{array} {lcl}
 & -ig^2[f_{abe}f_{cde}(g^{\mu \rho}g^{\nu \sigma}-g^{\mu \sigma}g^{\nu \rho})\\
 & +f_{ace}f_{bde}(g^{\mu \nu}g^{\rho \sigma}-g^{\mu \sigma}g^{\nu \rho})\\
 & +f_{ade}f_{bce}(g^{\mu \nu}g^{\rho \sigma}-g^{\mu \rho}g^{\nu \sigma})]\end{array}$  \\ [.1ex]
\\ [.3ex]
\hline\hline
\end{tabular}
\label{table:Fd}
\end{table}
\end{center}

\section{Loop integrals}
   The scalar loop integrals of the two-, three-, and four-point functions which are
used for the calculation of the self energy and the vertex diagrams are given by
\begin{align*}
&A_0(m^2)= \frac{(2\pi\mu)^{4-D}}{i\pi^2} \int d^Dk \frac{1}{k^2-m^2}, \\
&B_0(p_1^2, m_1^2, m_2^2)= \frac{(2\pi\mu)^{4-D}}{i\pi^2} \int d^Dk \frac{1}{[k^2-m_1^2][(k+p_1)^2-m_2^2]}, \\
&C_0(p_1^2, p_2^2, p_{12}^2, m_1^2, m_2^2, m_3^2)=\\
&\qquad\qquad \frac{(2\pi\mu)^{4-D}}{i\pi^2} \int d^Dk \frac{1}{[k^2-m_1^2][(k+p_1)^2-m_2^2][(k+p_1+p_2)^2-m_3^2},\\
&D_0(p_1^2, p_2^2, p_3^2, p_4^2, p_{12}^2, p_{23}^2, m_1^2, m_2^2, m_3^2, m_4^2)=\\
&\qquad\qquad \frac{(2\pi\mu)^{4-D}}{i\pi^2} \int d^Dk \frac{1}{[k^2-m_1^2][(k+p_1)^2-m_2^2],
[(k+p_1+p_2)^2-m_3^2][(k+p_4)^2-m_4^2]}
\end{align*}
with the abbreviation $p_{ij}=(p_i+p_j)$ for the momenta.
   For the sake of brevity, we have suppressed the boundary conditions $i\epsilon$ in the individual factors of
the denominators.

\end{appendix}

\end{document}